\documentstyle[12pt,epsfig]{article}

\large
\newcommand{\be}{\begin{eqnarray}}
\newcommand{\ee}{\end{eqnarray}}
\newcommand{\bea}{\left (\begin{array}{cc}}
\newcommand{\eea}{\right )\end{array}}

\newcommand{\mat}{\left ( \begin{array}{cc}}
\newcommand{\emat}{\end{array} \right )}
\newcommand{\matt}{\left ( \begin{array}{ccc}}
\newcommand{\ematt}{\end{array} \right )}
\newcommand{\matf}{\left ( \begin{array}{cccc}}
\newcommand{\ematf}{\end{array} \right )}
\newcommand{\vect}{\left ( \begin{array}{c}}
\newcommand{\evect}{\end{array} \right )}

\begin{document}
\setlength{\baselineskip}{17pt}
\pagestyle{empty}
\vfill
\eject
\begin{flushright}
SUNY-NTG-01/02
\end{flushright}

\vskip 2.0cm
\centerline{\Large \bf Spectral properties of a generalized chGUE}
\vskip 0.4cm
\centerline{\Large \bf }

\vskip 0.8cm
\centerline{A.M. Garc\'{\i}a-Garc\'{\i}a}
\vskip 0.2cm
\centerline{\it
Department of Physics and Astronomy, SUNY, 
Stony Brook, New York 11794}
\vskip 1.5cm

\centerline{\bf Abstract}
\noindent
We consider a  generalized 
chGUE based on a weak confining potential.  
We study the 
spectral correlations close to the origin in the thermodynamic limit.
 We show that
 for eigenvalues separated up to the mean 
level spacing the spectral correlations  coincide with those of chGUE. 
 Beyond 
this point, the spectrum is described by an oscillating  number variance 
centered
around a constant value. We argue that the origin of such 
a rigid spectrum is due to the breakdown of the translational invariance 
of the spectral kernel in the bulk of the spectrum. 
 Finally, we compare our 
results with the ones obtained from a critical chGUE recently reported 
in the literature.
We conclude that our generalized chGUE does not belong to the same 
class of universality as the above mentioned model.

\vfill

\eject
\pagestyle{plain}

\section{Introduction}
Critical statistics has been an intense subject 
of study in recent years \cite{chen,kra1,Moshe,dittes,log2,mirlin,nishigaki,pato1}. 
So far, two kinds of models have been proposed
to describe those critical correlations. In one of them, deviations from 
Wigner-Dyson statistics are obtained by  
adding a symmetry breaking term to the Gaussian Unitary Ensemble (GUE) {\cite{Moshe,us}}. 
The model is solved by mapping it to a non interacting Fermi gas of 
 eigenvalues. The second one {\cite{log2}} makes use of  soft confining potentials. 
It is solved exactly by means of q-orthogonal polynomials.

 Universality  in critical statistics
 have been conjectured \cite{kra1} due to the fact
  both models share the same kernel
in the thermodynamic limit and when the deviations
from GUE are small. However,
the origin of the critical kernel 
is different in both cases. In models based 
in soft confining potentials the 
critical kernel is obtained from a 
nontrivial unfolding caused by the 
strong fluctuations of the spectral density. In 
models with a explicit breaking symmetry term, deviations from Wigner-Dyson
statistics arise because the long range interactions among eigenvalues are
suppressed \cite{krav}.
  Although  progress have been recently reported \cite{tsvelik}, 
 the universality class associated with critical statistics can be
still considered an unresolved problem.  

Recently {\cite{us}}, a critical  chiral Gaussian Unitary Ensemble (chGUE) \cite{ver1} of the first class 
(addition of a 
symmetry breaking term to the chGUE) was proposed
in order to describe the spectral correlations of the QCD Dirac operator beyond
the Thouless energy \cite{ver}. It 
was found that in the bulk of the spectrum and for small deviations from 
the Wigner-Dyson statistics,  
the spectral kernel coincided with the one conjectured \cite{kra1}
 to be universal 
for critical statistics.
In this letter we shall 
 study the effect of the hard edge (the ensemble is defined on the 
positive real axis only) on the 
spectral correlations of a  chGUE with a weak confining potential.
 
We will proceed as follows.
 First, we propose a random matrix ensemble defined on the positive real line
 with a non-polynomial potential  
 which is soft confining in the bulk of the spectrum 
 and Gaussian close to the 
origin.
Then,  we compute  
the spectral kernel in the semiclassical approximation. Finally,
we compare our model with the above mentioned critical chGUE {\cite{us}.
 
Properties of chGUE with weakly confining potentials 
have already been discussed 
in the 
physics literature {\cite{ver2,chen,bleck,ism}}, but attention was 
focused in the bulk 
of the spectrum. The effect of the hard edge in the  critical spectral 
correlations and its impact on universality  
remains an open question.

\section{Definition of the Model}

In this section we introduce the model to be studied and argue the need to 
unfold the spectrum. Finally, we compute the mean spectral density 
needed for such unfolding by 
using the Dyson's mean field equation.

We consider a $N\times N$ complex hermitian matrix ensemble $H$ with 
block structure,
\be
H=\left(\begin{array}{cr} 0 & C^{\dagger} \\ C & 0\end{array}\right)
\ee 
and probability distribution 
 given by,
\be
\label{pot}
P(C) \propto e^{-V(CC^{\dagger})}
\ee
where $C$ is a $N/2 \times N/2$ 
hermitian matrix.
In terms of the eigenvalues of $H$, the joint distribution is given
by,
\be
\label{j}
P(x_{1} \ldots x_{N}) \propto \prod_{i=1}^{N}x_{i}e^{-V(x_{i})}
\prod_{1\leq i<j\leq N}|x_{i}^{2}
-x_{j}^{2}|^2
\ee
\be
\label{v}
V(x_i)=\frac{1}{\gamma}{\rm{arcsinh}}^2(x_i)
\ee
where $x_{i}$ are the eigenvalues of $H$.

 Since $V(x_i)$ in (\ref{j}) is proportional to $x_{i}^2$ for $x_{i} \ll 1$ 
we expect to recover the chGUE  kernel \cite{ver1}
in this limit. For $x_{i} \gg 1$,  the potential
 $V(x_{i}) \propto \log^{2}(x_i)$
fail to keep the eigenvalues confined and deviations from the chGUE
may be relevant \cite{bog}.

If the considered interval were the whole real line, the orthogonal 
polynomials associated with (\ref{v}) would be the $1/q$-
 Hermite polynomials $h_{n}(x;q)$ \cite{ask,atak3,log2}
 with $\gamma=\log(1/q)$. 
Unfortunately, for the positive real axes  
  we do not know
any  set of polynomials orthogonal with respect to 
the measure (\ref{pot}) with 
the potential (\ref{v}). Thus,
in order to compute the mean spectral density necessary to unfold the 
spectrum we shall use the Dyson's mean field equation. 

The joint distribution (\ref{j}) can be written as a statistical distribution of a 
one dimensional system of  $N$ particles  at 
temperature ``$1/T=2$'' with a  pairwise logarithmic interaction 
and a one particle potential given by (\ref{v}) that maintains the
system confined. 
\be
P(x_{1},\ldots x_N)\propto e^{-F(x_1, \ldots, x_N)/T}
\ee
where
\be
\label{s}
F(x_1, \ldots, x_N)=\sum^{N}_{i=1}V(x_{i})-2\sum_{i\neq j}^N\log|x_{i}^{2}-
x_{j}^{2}|-\log(x_{i})
\ee
We want to perform a mean field theory analysis of the above one dimensional
system. We assume that in the large $N$ limit the above system has a 
continuous macroscopic density given by, 
\be
\rho(\epsilon)=\sum_{i=1}^{N}\delta(\epsilon-
\epsilon_{i})
\ee 
Plugging $\rho(\epsilon)$ into (\ref{s}) and assuming
that the density is non zero only in the interval
$0<x<D$ we can express $F$ as a functional of 
the spectral density. 
\be 
F[\bar{\rho}(\epsilon)]=\int_{0}^{D}\bar{\rho}(\epsilon)V(\epsilon)d\epsilon
- \int_{0}^{D}\int_{0}^{D}
\bar{\rho}(\epsilon)\bar{\rho}(\epsilon')\log|\epsilon-\epsilon'|d\epsilon 
d\epsilon'
\ee

The mean spectral density $\rho_{MF}$ is defined as the 
density that minimizes the above functional, namely, $\delta F/\delta
\bar{\rho}=0$, that implies 
\be
\int_{0}^{D} d\epsilon'{\rho}(\epsilon_{MF}')\log|\epsilon-\epsilon'|=V(\epsilon)+c
\ee
where $c$ is a constant due to the normalization constraint.
The general solution of the above equation  (usually called Dyson equation)
is given by \cite{akh},

\be
\label{p}
\rho_{MF}(\epsilon)=\frac{1}{{\pi}^2}\sqrt{D^2-{\epsilon}^2}
Re\int_{0}^{D}\frac{dV/dt}{\sqrt{D^2-{t}^2}}\frac{tdt}{t^2-{\epsilon
_{+}^2}}
\ee
where $\epsilon_{+}=\epsilon +i0$.
$D$ is found from the normalization condition,
\be
\int_{0}^{D}\rho_{MF}(\epsilon)d\epsilon=N
\ee

Now,  the task is to compute $\rho_{MF}$ for the potential $V(\epsilon
)=\frac{1}{\gamma}{\rm{arcsinh}^{2}(\epsilon)}$,
\be
\label{m}
\rho_{MF}(\epsilon)=\frac{2}{\gamma{\pi}^2}\sqrt{D^2-{\epsilon}^2}
Re\int_{0}^{D}\frac{{\rm{arcsinh}}(t)}{\sqrt{1+t^2}}\frac{1}{\sqrt{D^2-{t}^2}}\frac{tdt}{t^2
-{\epsilon
_{+}^2}}
\ee
This integral can be performed by changing the contour of 
integration  in a sum 
of two pieces, $A=A_{1}+A_{2}$ where $A_1$ is the the negative imaginary
axis and $A_{2}$ is the interval $[D,\infty]$. Since we are interested only
in the real part of (\ref{m}), $A_{2}$ does not contribute to the integral
. Thus,
(\ref{m}) can be written as,
\be
\rho_{MF}(\epsilon)=\frac{1}{\gamma{\pi}}\sqrt{D^2-{\epsilon}^2}\frac{2}{\pi}\int_{1}
^{\infty}\frac{t}{\sqrt{t^2-1}\sqrt{D^2+t^2}}\frac{1}{t^2+{\epsilon}^2} 
\ee
The above integral can be performed by means of 
a change of variables, the final
result being,
\be
\rho_{MF}(\epsilon)=\frac{1}{\pi \gamma}\frac{1}{\sqrt{{\epsilon}^2+1}}\arctan
\frac{\sqrt{D^2-{\epsilon}^2}}{\sqrt{{\epsilon}^2+1}}
\ee
From the normalization condition we find that $D=\sinh(\pi \gamma N)$, therefore,
the mean spectral density for $N \rightarrow \infty$ is given by,
\be
\label{pol}
\rho_{MF}(\epsilon)=\frac{1}{2\gamma}\frac{1}{\sqrt{{\epsilon}^2+1}}
\ee
As expected, 
 $\rho_{MF}(\epsilon)$ has the right limiting values, it is a constant for 
$\epsilon \ll 1$ (as in the chGUE case)
and for $\epsilon \gg 1$ is proportional  to $1/\epsilon$, as for the 
random matrix ensemble  with soft confining potentials discussed in 
\cite{bog,nishigaki,ask,pato1}.
The above spectral density will be used in the next section to unfold the 
spectrum.

This unfolding allows us to  work in units in which the mean 
level spacing is equal to one.
 We recall that, in this context, random matrix theories only reproduce
spectral correlations around the average spectral density.
 
We remark that the above mean spectral density 
is an approximate formula capable of giving 
only the smooth part of the spectral density. 
The exact mean spectral density has 
oscillations which are out of reach 
of the mean field formalism above used. Therefore,
the mean spectral density (\ref{pol})  is only valid
approximation if these fluctuation are small enough \cite{nishigaki}. In our
model this situation corresponds with $\gamma \ll 1$. For $\gamma \gg 1$ the 
exact spectral density is a rapidly oscillating function. Hence, it is not 
possible to define a meaningful mean spectral density out of it
\cite{nishigaki,bog}.

As the mean spectral density 
is not constant, the rescaling procedure is not trivial {\cite{krav}}. 
The variable $x$ in terms
of which the spectral density becomes a constant, is the integrated 
mean spectral density.
\be
x=\int_{0}^{E}\rho_{MF}(\epsilon)d\epsilon
\label{unf}
\ee
where $\rho_{MF}(\epsilon)$ is  the mean spectral density previously found. We shall
see in the next section that this nontrivial unfolding is the main ingredient
to get a non translational invariant kernel in the bulk of the spectrum.

\section{Calculation of the spectral kernel}

 In this section we compute the spectral kernel in the semiclassical approximation.
The semiclassical approximation in the GUE consists in substituting the wave function
(orthogonal polynomials times $e^{-V(x)}$) appearing in the spectral kernel, 
after the Christoffel-Darboux formula is applied, by their WKB approximation. Because of
the hard edge at $x=0$ we cannot simply do a WKB approximation by replacing the wave functions by 
plane waves, but instead have to use Bessel functions.
 \begin{figure}[ht]  
\begin{center}
\epsfig{figure=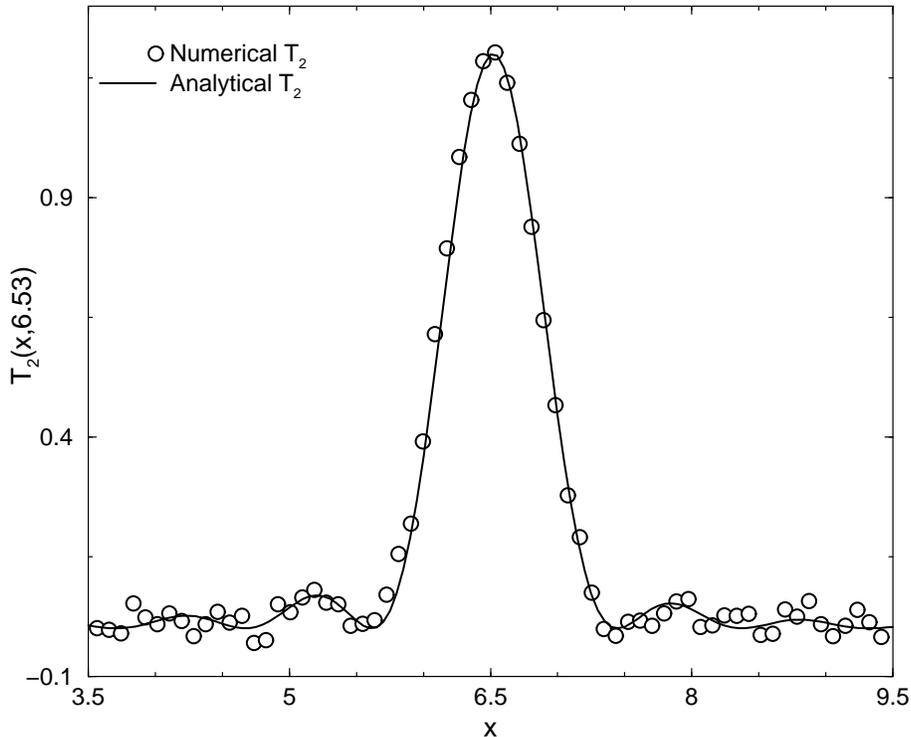,width=100mm,angle=270}
\caption{We compare the analytical
value of $T_{2}(x,6.53)=|K(x,6.53)|^{2}$  with the numerical one.
$K(x,y)$ is given by (\ref{ker}) with $\gamma=0.25$.
The numerical integration was performed by using the Metropolis 
algorithm for $N=100$ ``particles'' in the confining potential (\ref{v}). We 
have used  the first $9{\cdot}10^4$ sweeps to ``warm up'' the system  and taken
 the average over the next $9{\cdot}10^{5}$. We repeated the process four times.
The final result is the the average of the four trials. We explicitly checked
the agreement between numerical and analytical results for 
$y<20$.}
\end{center}
\end{figure}      
The kernel associated with
(\ref{v}) can be written in terms of the wave function as follows,  
\be
\label{ke}
K(E,E')=\frac{\psi_{2N+1}(E)\psi_{2N}(E')-\psi_{2N+1}
(E')\psi_{2N}(E)}{\pi(E-E')}+\frac{\psi_{2N+1}(E')\psi_{2N}(E)+\psi_{2N+1}
(E')\psi_{2N}(E)}{\pi(E+E')}
\ee
where $\psi_{2N}(E)$ and $\psi_{2N+1}(E)$ are the even and odd large $N$ limit 
of the wave functions associated with the potential (\ref{v}). 
In the semiclassical approximation those functions 
are given by \cite{bog},
\be
\label{puteo}
\psi_{2N}(E)=\sqrt{s(E)}J_{0}(\pi s(E)) \\ \nonumber
\psi_{2N+1}(E)=\sqrt{s(E)}J_{1}(\pi s(E))
\ee
where $J_{0}$ and $J_{1}$ are Bessel functions and $s(E)$ is defined
by, $\rho_{MF}=ds/dE$, with  $\rho_{MF}$ the mean spectral density 
computed in the last section. 
It is clear \cite{nigel} that the above semiclassical
expressions for the wave functions are correct for polynomial increasing
potential. For soft confining potentials, according to \cite{nishigaki,bog}
these expressions are valid if the mean-field spectral density used to unfold
the spectrum is close to the exact mean spectral density. In our model 
this happens whether $\gamma \ll 1$.
Othe argument supporting (\ref{puteo}) comes from the asymptotic
form of the orthogonal polynomials associated with a potential asymptotically
proportional to $\log^2(E)$. This problem have already been discussed in the 
literature \cite{chen}. They found that for $E \gg 1$, 
$\Psi_{2N}(E) \propto \cos(\log(E)/\gamma)$. This result coincides with 
(\ref{puteo}) in the limit considered. For $E \ll 1$, $s(E) \propto 1/\gamma$
and we recover the chGUE result. As an additional check, 
we evaluate $T_{2}(x,y)=|K(x,y)|^{2}$ by numerical integration of the
the joint distribution (\ref{j}). Figure 1 shows that the agreement 
between numerical and analytical results is excellent.           
We recall that for polynomial-like increasing
potentials the mean spectral density is a constant proportional to $N \gg 1$, 
$s(E)$ is linear for $E \rightarrow 0$ and we recover 
the kernel of the chGUE \cite{ver2}. 

Once we know the form of the kernel  we can unfold the spectrum by using
(\ref{unf}) and (\ref{pol}) \cite{krav,bleck}.

\be
\int_{0}^{E}\frac{2}{\gamma}\frac{1}{\sqrt{1+{\epsilon}^2}}d\epsilon=x
\ee
\be
E=\sinh(x\gamma/2)
\ee  
where for convenience we have set $\gamma=\gamma/4$.
The kernel in terms of the new, unfolded variables is given by,
\be
K(x,y)=\frac{\pi \gamma}{8}\sqrt{\cosh(\gamma x/2)\cosh(\gamma y/2)xy}
[\frac{J_{0}(\pi x)J_{1}(\pi y)+J_{0}(\pi y)J_{1}(\pi x)}{\sinh((x+y)
\gamma/4)\cosh((x-y)\gamma/4)}+\nonumber \\
\frac{J_{1}(\pi x)J_{0}(\pi y)-
J_{1}(\pi y)J_{0}(\pi x)}{\sinh((x-y)\gamma/4)\cosh((x+y)\gamma/4)}]
\label{ker}
\ee
As expected, for $\gamma \rightarrow 0$ we recover the chGUE kernel.  
This kernel is already in a suitable form for comparison with the one 
previously found in 
 {\cite{us}}.
\be 
\label{kerus}
K(x,y)=\frac{\pi\gamma}{8}\sqrt{xy}[\frac{J_{0}(\pi x)J_{1}(\pi y)+
J_{0}(\pi y)J_{1}(\pi x)}{\sinh((x+y)\gamma /4)}+\frac{J_{1}(\pi x)
J_{0}(\pi y)-J_{1}(\pi y)J_{0}(\pi x)}{\sinh((x-y)\gamma/4)}]
\ee

\begin{figure}[ht]  
\begin{center}
\epsfig{figure=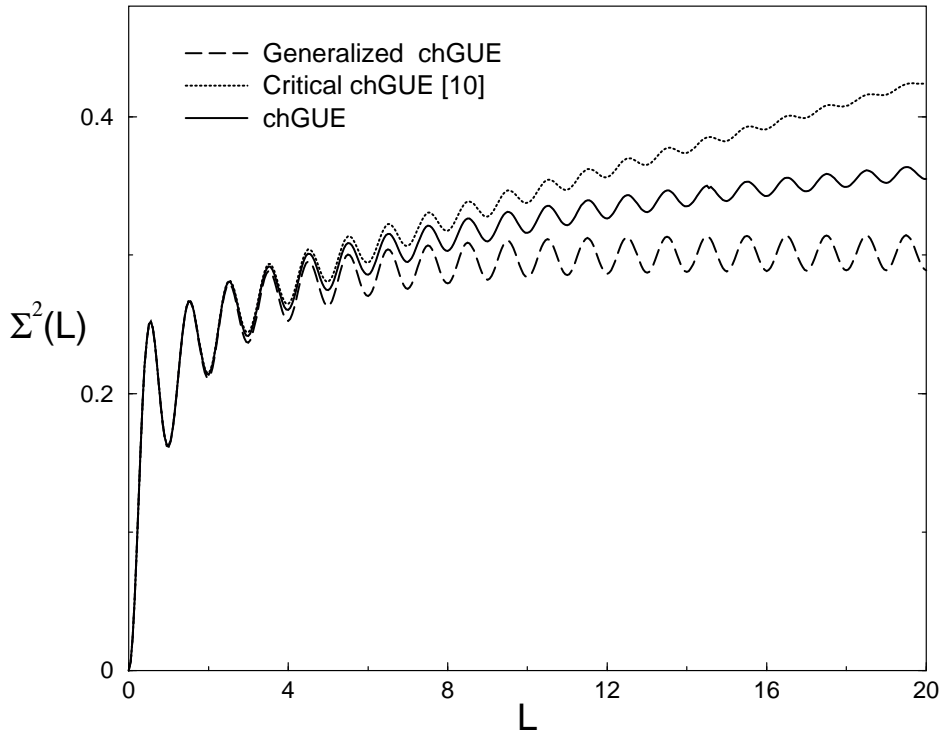,width=100mm,angle=270}
\caption{Number variance for $\gamma=\log(1/q)=0.1$. It is linear for $L \gg 1$
in the critical chGUE of \cite{us}. In our generalized chGUE is almost constant in the 
bulk of the spectrum.}
\end{center}
\end{figure} 
Even though both kernels reproduce the chGUE kernel 
for $\gamma \rightarrow 0$ they are essentially different in the bulk of the 
spectrum. (\ref{kerus}) is translational invariant in the 
bulk of the spectrum, unlike (\ref{ker}), which is not. 
 The origin of such non 
traslationally invariant kernel
is due to the nontrivial unfolding induced by 
the mean spectral density. This unfolding prevents from 
vanishing the second term of the right hand of (\ref{ke}) in the bulk of
the spectrum.

In the next section, we shall
study the effect of the non-translational 
invariance of the kernel in the spectral correlations involving many levels 
by computing the number variance.

\section{Discussion of results}
In this section we shall see, by computing the number variance, that the 
spectrum of our model is more rigid than the chGUE one and 
essentially different
from the models describing critical statistics.
 
 In order
to  observe deviations from chGUE prediction  we are
going to study long-range correlations of eigenvalues 
by studying the number variance in an interval $[0,s]$.\\ The number 
variance is a statistical quantity which gives a quantitative description 
of the stiffness of the spectrum. The number variance 
is obtained by
integrating the two-point correlation function including the self-correlations
\be
\Sigma^2(L) = \int_0^{L} dx\int_0^{L} dy\left [
\delta(x-y) \langle \rho(x) \rangle + R_2(x,y) \right ].
\ee   
\begin{figure}[ht]  
\begin{center}
\epsfig{figure=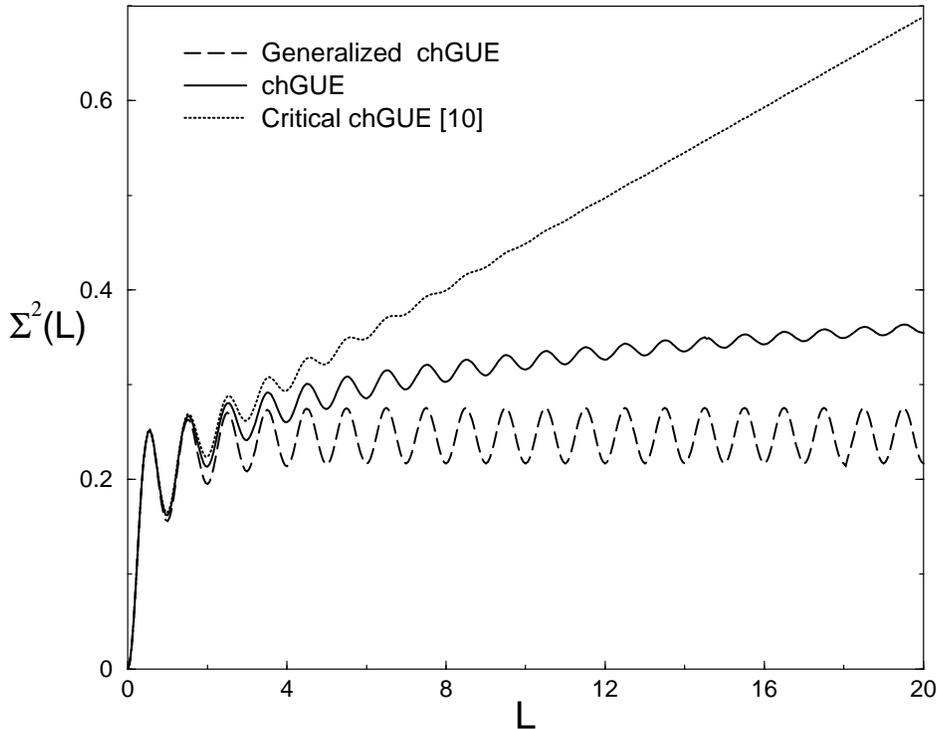,width=100mm,angle=270}
\caption{$\gamma=\log(1/q)=0.3$. A constant number variance is 
observed in the generalized chGUE 
when $L\gg 1$. That means the spectrum is even more rigid
than the chGUE one.}
\end{center}
\end{figure}  
For the  Wigner-Dyson statistics the 
number variance is  proportional to $\log(L)$. Such weakly
increasing number variance is not surprising as the eigenvalues repulsion 
produces a highly rigid spectrum.    
 For the Poisson statistics the 
number variance is equal to $L$ as expected for eigenvalues which
are not correlated.
Finally, a number variance proportional to $\chi L$ (for $L \gg 1$ and 
$\chi \ll 1$) 
is a signature of
critical statistics {\cite{aronov,braun,bleck,mirlin,kra2,nishigaki}}. The  
slope $\chi$  is directly related {\cite{chalker2}} to the multifractality
of the wave functions of a disordered system at a 
delocalization-localization transition.

 A  linear number variance for $L \gg 1$ with a slope $\chi \ll 1$ was found in 
the generalized chGUE {\cite{us}}. 
However, as it can be observed in Figure 2 and 3,
the number variance of our model is almost constant for $L \gg 1$. 
The oscillating behavior around a constant value is 
partially due to the self-interactions
coming from the first term of the number variance.

Apparently, this result is surprising because  random matrix ensembles with 
broken time invariance  
based on  potentials behaving
as $\log^{2}(x)$ asymptotically are supposed to have a linear 
number variance with 
slope $\chi$ in the bulk of the spectrum, which is a
signature of critical statistics \cite{nishigaki,log2}.
 In principle, one may think that the
presence of a hard edge at $x=0$  in 
our model does not affect the spectral properties in the bulk of the spectrum.
We argue that this is not the case. 

The hard edge, combined with the 
soft-confining 
nature of the potential breaks up the translational invariance 
of the kernel  (\ref{ker}) even in the bulk of the spectrum.
In the bulk , the cluster function associated with the 
kernel (\ref{ker})is 
given by,
\be
\label{pl}
Y_{2}(x,y) \propto [\frac{\sin^2(\pi(x-y))}{\sinh^2((x-y)\gamma/4)}+
\frac{\sin^2(\pi(x+y))}{\cosh^2((x-y)\gamma/4)}+
2\frac{\sin(\pi(x+y))}{\cosh((x-y)\gamma/4)}\frac{\sin(\pi(x-y))}
{\sinh((x-y)\gamma/4)}]
\ee 
where we have used the asymptotic expression of the Bessel functions.
By performing  elementary integrations, we observe 
that the leading contribution to the number variance for $L \gg 1$
 coming from  the first term  of $Y_{2}$ (translational
invariant part) is $\chi L$ where $\chi$ is a function of $\gamma$ only. On the
other hand, the leading contribution of the second term of 
$Y_{2}$ (non-translational
invariant part) to the number variance is $-\chi L$.   
  Therefore, both contributions cancel each other and we are left with 
a oscillating (around a constant value depending on $\gamma$) number
variance coming from the third term of the cluster function.
We point out that the above cancelation is mainly due to the non-trivial 
unfolding used. 

Roughly speaking, weak increasing potentials fail to keep 
the eigenvalues confined. As a consequence, the mean spectral density is
in general a strongly oscillating function even in the thermodynamic limit. If 
$\gamma \ll 1$ the deviations from chGUE are 
small and we can still define a relevant smooth 
mean spectral density by using the mean field formalism \cite{bog}. 
The unfolding
procedure using this mean spectral density
 breaks the translational invariance of the kernel in the bulk of the 
spectrum. This breaking of the translational symmetry produces a spectrum
highly correlated and essentially different from the one reported in {\cite{us}}.

 We would like to mention that a similar result to the one found in this
letter has been reported by Canali and Kravtsov {\cite{canali,krav}} .
They studied the spectral properties of a  generalized  GUE 
based on a weak confining potential with a $\log^2(x)$ asymptotic as well.
 They noticed
 that  in the bulk of the spectrum for $N \rightarrow \infty$
and $\gamma <<1$, the cluster function $Y_{2}(x,y)$ 
of  that ensemble has strong correlations not only when
$x \approx y$, but also when $x \approx -y$. The total 
cluster function is given by the following non translational invariant
relation,
\be 
Y_{2}(x,y)=\frac{\gamma^2}{16{\pi}^2}\frac{{\sin}^{2}
(\pi(x-y))\cosh(x\gamma/2)\cosh(y\gamma/2)}
{\cosh^2((x+y)\gamma/4)\sinh^2((x-y)\gamma/4)}
\ee

They showed that, due to this `ghost' peak,  the number variance 
depends on the interval in which it is calculated.

If the interval is not symmetric with respect to the origin ($[0,s]$ for
instance), the 
system does not feel the strong (non-translational invariant)
 correlations at $x=-y$. Then, the number 
variance goes  asymptotically like  $\chi L$ and the model
is supposed to describe critical correlations.
However, if the interval is symmetric with respect to the origin , the peak
at $x=-y$ of the two point function $Y_{2}(x,y)$ has to be taken into account as
well. This contribution  drives the asymptotic form of the number variance 
to a constant value {\cite{kra2,canali,wallin}}, in agreement
with the results obtained in this letter.

It is straightforward to connect the number variance in the bulk 
of the spectrum of {\cite{kra2,canali}} with the one studied in this letter.
The asymptotic form of the number variance in a interval $[0,s]$
associated with the  first two terms of (\ref{pl})  
corresponds to the number variance in the interval $[-s/2,s/2]$ 
of the  above mentioned critical GUE. By changing variables $u=-x$,$v=y$ 
in the expression for the number variance of our model we recover 
the expression obtained by Canali and Kravtsov.
 The third term of (\ref{pl}) produces the oscillating
behavior observed only in our generalized chGUE.

 To sum up,
 due to the non translational
invariance of the kernel 
 contributions coming from the points $x \sim -y$
 have to be taken into account. 
These contributions  make the linear term 
in the number variance vanish. 

\section{Conclusions}

In this letter we have studied the effect of a hard edge
in the spectral correlations of  
a  chiral random matrix ensemble with a soft confining potential. 
We showed that beyond the Thouless energy the
spectrum is characterized by an oscillating 
number variance around a constant value. The spectrum is even more correlated  
than the chGUE one.

The linear term of the number variance characterizing 
critical statistics vanishes due to the non-translational invariance of the 
spectral kernel in the bulk of the spectrum.  
Thus, the generalized chGUE studied in this letter  
and the 
ensemble of  {\cite{us}} (in which a  linear number variance
was found to be proportional to $\chi L$ for $L>>1$) belong   
to different universality classes {\cite{wallin}}.

 \vskip 0.5 cm
{\bf Acknowledgements}
I thank  Daniel Robles, Prof. Denis Dalmazi, Prof. Y.Chen 
 for important suggestions and useful 
discussions. I thank Prof. J.J.M Verbaarschot for illuminating discussions and 
for a critical reading of the manuscript. I am indebted to James Osborn 
for providing me the program to perform the numerical simulation.
\vskip 0.5cm
This work was supported by the US DOE grant
DE-FG-88ER40388

\end{document}